\documentclass[%
 reprint,
 amsmath,amssymb,
aps,
prfluids,
onecolumn,
longbibliography,
superscriptaddress,
notitlepage,
floatfix
]{revtex4-2}

\usepackage{graphicx}
\usepackage{dcolumn}
\usepackage{bm}

\usepackage[usenames]{color}

\begin{document}


\title{
Data-driven prediction
of reversal of large-scale circulation in turbulent convection}

\author{Daigaku Katsumi}%
\affiliation{%
Department of Mechanical Engineering, Tokyo Denki University, Adachi, Tokyo 120-8551 Japan
}%

\author{Masanobu Inubushi}
\affiliation{
Department of Applied Mathematics, Tokyo University of Science, Shinjuku, Tokyo 162-8601 Japan
}%

\author{Naoto Yokoyama}
\email{n.yokoyama@mail.dendai.ac.jp}
\affiliation{%
Department of Mechanical Engineering, Tokyo Denki University, Adachi, Tokyo 120-8551 Japan
}%

\date{\today}

\begin{abstract}
Large-scale circulation (LSC) quasi-stably emerges in the turbulent Rayleigh-B\'{e}nard convection,
and intermittently reverses its rotational direction in two-dimensional turbulent convection.
In this paper,
direct numerical simulations of the intermittent reversals of the LSC
in a two-dimensional square domain are performed,
and the time series of the total angular momentum
indicating the rotational direction of the LSC
is predicted by reservoir computing
whose input consists of the shear rates and temperatures at six locations on the sidewalls.
The total angular momentum in the simulation after times
shorter than half the typical duration of the quasi-stable states
is successfully reproduced by the locally-measurable quantities on the sidewalls
because the secondary rolls accompanied by the boundary flow characterize the reversal of the LSC.
The successful prediction by such sparse input derived from local measurements on the sidewalls demonstrates that
the reservoir computing prediction of the reversal is feasible in laboratory experiments and industrial applications.
On the other hand,
long-term prediction often provides
the total angular momentum opposite in sign to the one in the simulations
in the late parts of long quasi-stable states.
The statistical independence of each reversal
implies that the prediction after the reversal is difficult or even impossible,
and
the training data in the late part in the long quasi-stable state,
which rarely appears,
is contaminated by the statistically-independent angular momentum in the subsequent quasi-stable state.
\end{abstract}

\maketitle

\section{Introduction}
\label{sec:introduction}

In Rayleigh-B\'enard systems where a fluid is heated below and cooled from above,
the flow regimes change from conduction to convection
as the Rayleigh number increases~\cite{chandrasekhar1981hydrodynamic,drazin_reid_2004}.
The Rayleigh-B\'enard convection (RBC) has been widely investigated
from the scientific and industrial viewpoints.
(See Ref.~\cite{RevModPhys.81.503,doi:10.1146/annurev.fluid.010908.165152,doi:10.1146/annurev.fluid.32.1.709,doi:10.1146/annurev.fl.26.010194.001033} and reference there in.)
In RBC,
quasi-stable large-scale circulations (LSCs) often emerge
depending on the container shape~\cite{PhysRevFluids.6.090502},
resulting from convection of small plumes~\cite{doi:10.1063/1.166473,PhysRevLett.75.4618}.
The LSCs play a crucial role in heat transport~\cite{doi:10.1063/5.0024408}.
Machine learning can successfully predict the Nusselt number and the Reynolds number
which respectively characterize the heat transport and the turbulence
of the turbulent statistics of RBC~\cite{doi:10.1063/5.0083943}.

In two-dimensional (2D) turbulent RBC,
the LSC intermittently reverses its rotational direction,
and the inversely rotating LSC emerges as another quasi-stable state~\cite{PhysRevE.83.067303,PhysRevE.84.026309,PhysRevFluids.6.033502}.
Such reversals were also observed in quasi-2D experiments~\cite{ni_huang_xia_2015}.
In the three-dimensional cube,
the LSC undergoes intermittent reorientations between the two diagonal planes of the cube~\cite{PhysRevE.95.033107,doi:10.1063/5.0021667}.
Similar reversals of the LSC in the turbulent background often appear in dynamo,
which is a model of the geomagnetical reversal~\cite{Glatzmaiers1995,Fauve_2017,Berhanu_2007,10.1093/gji/ggu287}.
The intermittent reversals of the primary LSC
in quasi-2D square container with solid boundaries
were observed for a wide range of the Rayleigh number and the Prandtl number,
and are considered to be triggered by growth of smaller counter-rotating secondary rolls near diagonally opposing two corners~\cite{PhysRevLett.105.034503}.
The reversals can be suppressed and enhanced by statically controlling the secondary rolls~\cite{zhang_xia_zhou_chen_2020,zhang_chen_xia_xi_zhou_chen_2021,doi:10.1080/14685248.2021.1916023,zhao_wang_wu_chong_zhou_2022}.

Model reduction to extract essential features is required
for dynamic control of turbulence~\cite{doi:10.1146/annurev-fluid-010816-060042,jones_heins_kerrigan_morrison_sharma_2015}.
A proper orthogonal decomposition (POD) revealed that
one of the energetic modes,
which detaches the primary LSC from the boundary layers before the reversals,
can be used as a precursor of the reversals~\cite{PhysRevE.95.013112}.
Although POD is useful for model reduction,
it is difficult to obtain the amplitudes of active modes at every moment.
Reconstruction of quantities of interest from a limited number of measurements
is usually needed in experiments and industrial applications~\cite{PhysRevFluids.6.100501,8361090}.
Furthermore,
the reconstruction of inner flows in wall-bounded turbulent flows from the stresses on the wall~\cite{10.1063/1.5128053,10.1063/5.0058346}
is expected to be used for closed-loop control based on non-intrusive sensing.

It is difficult to predict the intermittent reversals of the LSCs,
because the transition from the quasi-stable state having a LSC
to the other quasi-stable state having the opposite LSC
is completed in times much shorter than the inter-reversal times,
i.e., the duration of the quasi-stable states.
Machine learning is one of the most promising approaches
for the prediction of such chaotic dynamics,
and has been intensively investigated~\cite{RevModPhys.91.045002,doi:10.1146/annurev-fluid-010719-060214}.
Among many machine learning methods for model-free prediction,
reservoir computing (RC) requires the minimal computing resources~\cite{PhysRevLett.120.024102,PhysRevResearch.2.012080},
and small amount of learning even for turbulence~\cite{PhysRevE.98.023111,PhysRevE.102.043301}.
RC is also advantageous for pattern recognition of time series.
The large-scale evolution and low-order statistics of a two-dimensional turbulent convection flow
was successfully reproduced by a RC model
trained by time evolution of the first 150 POD modes~\cite{PhysRevFluids.5.113506}.
Extreme events of vorticity observed in experiments were also
reproduced by a RC model~\cite{PhysRevResearch.4.023180}.

In this paper,
direct numerical simulations (DNS) of 2D turbulent RBC are performed,
and time series of shear rates and temperatures at six locations on the sidewalls
are obtained.
RC with such sparse input derived from the local measurements on the sidewalls
is employed to predict the time series of the angular momentum and hence the reversals of the LSC.
The paper is organized as follows.
The numerical methods for DNS of RBC,
and for RC prediction of the time series of the angular momentum
are introduced in Sec.~\ref{sec:numericalmethod}.
The results of the DNS and the RC prediction are demonstrated in Sec.~\ref{sec:results}.
The selection of the input to the RC
and the difficulty of the long-term prediction
are discussed from the viewpoint of the flow characteristics
in Sec.~\ref{sec:discussion}.
The last section is devoted for conclusion.

\section{Numerical Method}
\label{sec:numericalmethod}

Under the Oberbeck-Boussinesq approximation,
the governing equation of RBC in a non-dimensional form
for velocity $\bm{u}$ and temperature $\theta$
as well as pressure $p$ is written as
\begin{subequations}
\begin{eqnarray}
 &&
 \nabla \cdot \bm{u} = 0
 ,
 \\
 &&
 \frac{\partial \bm{u}}{\partial t} + (\bm{u} \cdot \nabla) \bm{u}
 = - \nabla p + \theta \bm{e}_z + \sqrt{\frac{Pr}{Ra}} \nabla^2 \bm{u}
 ,
 \\
 &&
 \frac{\partial \theta}{\partial t} + (\bm{u} \cdot \nabla) \theta
 = \frac{1}{\sqrt{Ra Pr}} \nabla^2 \theta
,
\end{eqnarray}
\end{subequations}
where the velocity, the time and the temperature
are scaled by the free-fall velocity, the free-fall time and the temperature difference between the bottom and the top, respectively.
The non-dimensional parameters,
the Rayleigh number $Ra$ and the Prandtl number $Pr$,
must be appropriately selected.
At inappropriate $Ra$ and $Pr$,
the LSCs do not emerge,
the secondary rolls and hence the reversals are suppressed,
or the duration of the quasi-stable states is too long (See Ref.~\cite{PhysRevLett.105.034503}.).
In this paper,
$Ra=1 \times 10^8$ and $Pr=4$ are employed
in order that the intermittent reversals occur in reasonable computational times.

\begin{figure}
\includegraphics[width=.9\textwidth]{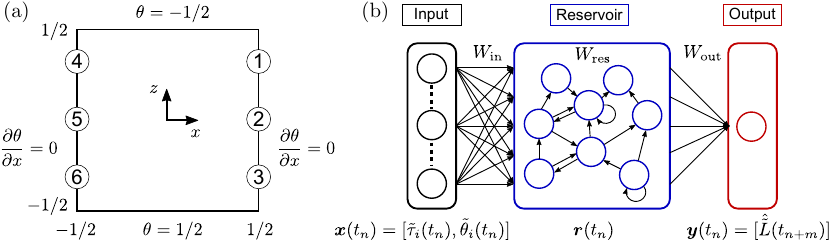}
\caption{
(a) Numerical domain.
The circled numbers represent the locations to measure $\dot{\gamma}_i$ and $\theta_i$.
(b) Algorithm of echo state networks with leaky-integrator neurons.
}
\label{fig:nd_alg}
\end{figure}

The horizontal and vertical coordinates are respectively denoted by $x$ and $z$,
and
the numerical domain is set to be $[-1/2,1/2] \times [-1/2,1/2]$.
The no-slip boundary condition for the velocity is applied at all the walls.
The temperature at the bottom, $z=-1/2$, and that at the top, $z=1/2$,
 are respectively set to be $1/2$ and $-1/2$.
The adiabatic boundary condition for the temperature $\partial \theta/\partial x=0$
is applied at the sidewalls $x=\pm 1/2$.
The numerical domain is shown in Fig.~\ref{fig:nd_alg}(a).

The Chebyshev collocation method
and the $\mathbb{P}_N$--$\mathbb{P}_{N-2}$ projection method with the third-order Adams-Bashforth backward-differentiation scheme
are used respectively for the spatial discretization and the time integration~\cite{peyret2002spectral}.
In this paper,
the number of the Gauss-Lobatto points is $N_{\mathrm{GL}} = 257 \times 257$,
and the time step is $5 \times 10^{-4}$.

The algorithm for prediction in the present study is
echo state networks with leaky-integrator neurons (LI-ESN),
which is a type of RC~\cite{JAEGER2007335},
and schematically drawn in Fig.~\ref{fig:nd_alg}(b).
The input vector, the reservoir vector and the output vector at $t_n$
are respectively represented by $\bm{x}(t_n) \in \mathbb{R}^{N_{\mathrm{in}}}$,
$\bm{r}(t_n) \in \mathbb{R}^{N_{\mathrm{res}}}$,
and $\bm{y}(t_n) \in \mathbb{R}^{N_{\mathrm{out}}}$.
Here, $t_n$ represents the discrete time of the RC,
and $N_{\mathrm{in}}$, $N_{\mathrm{res}}$, and $N_{\mathrm{out}}$ denote the numbers of nodes at each layer.
The evolution of $\bm{r}(t_n)$
and $\bm{y}(t_{n})$ are given as
\begin{subequations}
\begin{eqnarray}
&&
  \bm{r}(t_{n+1}) = (1-\alpha) \bm{r}(t_{n})
  + \alpha \bm{f}(W_{\mathrm{in}} \bm{x}(t_{n+1}) + W_{\mathrm{res}} \bm{r}(t_{n}))
  ,
  \\
&&
  \bm{y}(t_{n}) = W_{\mathrm{out}} \bm{r}(t_{n})
  ,
\end{eqnarray}
\end{subequations}
where $N_{\mathrm{res}}=1000$ is used.
In this paper,
the elements of the input weight matrix $W_{\mathrm{in}} \in \mathbb{R}^{N_{\mathrm{res}} \times N_{\mathrm{in}}}$ have a uniform distribution.
The connectivity of the reservoir weight matrix $W_{\mathrm{res}} \in \mathbb{R}^{N_{\mathrm{res}} \times N_{\mathrm{res}}}$ is fixed at $C_{\mathrm{res}}=0.05$,
each non-zero element is randomly drawn from a uniform distribution,
and the spectral radius of $W_{\mathrm{res}}$ is set to $\rho = 0.7$.
The output weight matrix $W_{\mathrm{out}} \in \mathbb{R}^{N_{\mathrm{out}} \times N_{\mathrm{res}}}$ is determined
in the training phase below.
The leak ratio $\alpha$ is set to be $0.3$,
and the element-wise hyperbolic tangent is used for the activation function $\bm{f}$.
The selection of these hyper-parameter values are provided in Appendix~\ref{sec:hyperparameters}.

The total angular momentum $L=\int (z u_x - x u_z) dV$
indicates the direction of the LSC:
$L>0$ and $L<0$ respectively indicate a clockwise rotation and a counter-clockwise one.
The reversals correspond to the sign inversion of $L$,
and the present objective is prediction of the sign inversion.
As shown in Ref.~\cite{PhysRevE.95.013112} and Sec.~\ref{sec:results} in this paper,
POD demonstrates that
the development of the secondary rolls that triggers the reversals
can be captured by two energetic modes,
and the spatial eigenfunctions of the two modes
have characteristic structures near the sidewalls.
Such particularity of the two modes indicates that
time series of shear rates $\dot{\gamma} = \partial u_z/\partial x$ and temperatures $\theta$
at several locations on the sidewalls
can characterize the reversals.
In this paper,
$\dot{\gamma}$ and $\theta$ at the six locations,
$x=\pm 1/2$, $z=\pm \sqrt{2}/4, 0$ (Fig.~\ref{fig:nd_alg}(a)),
obtained by the DNS of RBC
are employed as the input vector,
i.e.,
 $\bm{x}(t_{n}) = [\tilde{\dot{\gamma}}_i(t_n), \tilde{\theta}_i(t_n)]$
($i=1,\ldots,6$),
and $N_{\mathrm{in}}=12$.
The selection of the locations will be discussed in
Sec.~\ref{sec:discussion}.
The input vector is normalized to the range $[-1,1]$ expressed by the tilde,
e.g., $\tilde{\dot{\gamma}}_i(t_n) = \dot{\gamma}_i(t_n) / \dot{\gamma}_{\mathrm{max}}$,
where $\dot{\gamma}_{\mathrm{max}}$ is the maximal shear rate over the six locations and time.
Moreover,
the one-sided linear weighted moving average with 10 times is applied to smooth each element of the input vector.

The output vector is prediction of the total angular momentum after $m$ steps of the RC,
i.e., $\bm{y}(t_{n}) = [\hat{\tilde{L}}_{m \delta}(t_{n+m})]$,
where the hat $\hat{\cdot}$ represents the prediction value,
and $N_{\mathrm{out}}=1$.
Because the time step for RC, $\delta$, corresponds to $0.2$ time units in the numerical simulation of RBC,
the RC predicts the total angular momentum after prediction lead time $T_{\mathrm{lead}} = m \delta$.
The prediction lead time is referred to simply as the lead time from now on.
Namely,
the RC predicts
the total angular momentum at $t_{n+m} = t_n + m \delta = t_n + T_{\mathrm{lead}}$
from the shear rates and temperatures on the sidewalls at $t_n$.
For example,
$\hat{\tilde{L}}_{0}(t_n)$, $\hat{\tilde{L}}_{0.2}(t_n+0.2)$, and $\hat{\tilde{L}}_{200}(t_n+200)$,
are predicted from $\tilde{\dot{\gamma}}_i(t_n)$ and $\tilde{\theta}_i(t_n)$
when $m=0$, $m=1$, and $m=1000$, respectively.
Note that $m=0$ ($T_{\mathrm{lead}}=0$) corresponds to simultaneous prediction, i.e., inference,
while $m=1$ ($T_{\mathrm{lead}}=0.2$) and $m \geq 2$ ($T_{\mathrm{lead}} \geq 0.4$) correspond to single-step-ahead and multi-step-ahead predictions, respectively.

The training data consist of
$\bm{x}(t_{n}) = [\tilde{\dot{\gamma}}_i(t_{n}), \tilde{\theta}_i(t_{n})]$
and $\bm{d}(t_{n}) = [\tilde{L}(t_{n+m})]$,
where $-T_{\mathrm{train}}-T_{\mathrm{lead}} \leq t_n \leq -T_{\mathrm{lead}}$,
$-T_{\mathrm{train}} \leq t_{n+m} \leq 0$.
The training length $T_{\mathrm{train}} = N_{\mathrm{train}} \delta$ is set to be $4 \times 10^4$.
Then,
the ridge regression determines the optimal output weight matrix
as
\begin{eqnarray}
W_{\mathrm{out}}^{\mathrm{opt}} = D R^{\mathrm{T}} (R R^{\mathrm{T}} + \beta I)^{-1}
\label{eq:Woutopt}
\end{eqnarray}
that minimizes
\begin{eqnarray}
\|D-W_{\mathrm{out}} R\|_{\mathrm{F}}^2 + \beta \|W_{\mathrm{out}}\|_{\mathrm{F}}^2
,
\end{eqnarray}
where
$R = [\bm{r}(t_1), \bm{r}(t_2), \ldots, \bm{r}(t_{N_{\mathrm{train}}})] \in \mathbb{R}^{N_{\mathrm{res}} \times N_{\mathrm{train}}}$,
$D = [\bm{d}(t_1), \bm{d}(t_2), \ldots, \bm{d}(t_{N_{\mathrm{train}}})] \in \mathbb{R}^{N_{\mathrm{out}} \times N_{\mathrm{train}}}$,
$I$ is the $N_{\mathrm{res}} \times N_{\mathrm{res}}$ identity matrix,
and $\|\cdot\|_{\mathrm{F}}$ denotes the Frobenius norm of a matrix.
The regularization parameter $\beta$ is set to be $5 \times 10^{-4}$ in this paper.

In the testing phase for time interval $T_{\mathrm{test}} = 2 \times 10^4$,
by using $W_{\mathrm{out}}^{\mathrm{opt}}$ obtained in the training phase,
$\hat{\tilde{L}}(t_{n+m})$ where $0 \leq t_{n+m} \leq T_{\mathrm{test}}$
is predicted from $\tilde{\dot{\gamma}}_i(t_{n})$ and $\tilde{\theta}_i(t_{n})$
where $-T_{\mathrm{lead}} \leq t_n \leq T_{\mathrm{test}}-T_{\mathrm{lead}}$.

\section{Results}
\label{sec:results}

\begin{figure}
\includegraphics[width=.95\textwidth]{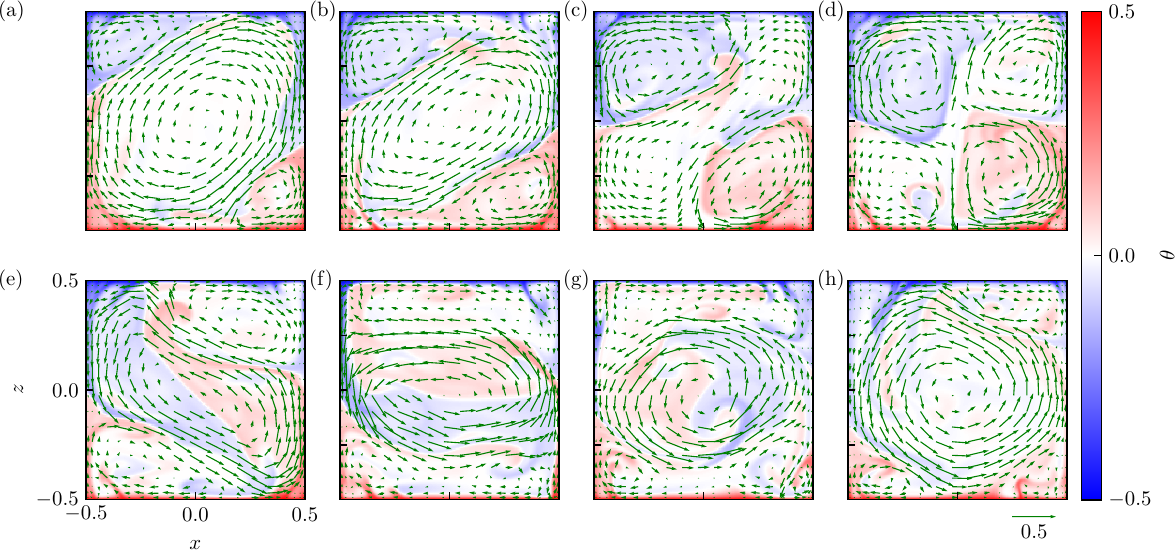}
\caption{
Flow field during a reversal.
The arrows and the color map respectively represent
the velocity and the temperature.
(a) $t=460$,
(b) $t=490$,
(c) $t=506$,
(d) $t=509$,
(e) $t=512$,
(f) $t=514$,
(g) $t=516$, and
(h) $t=540$.
}
\label{fig:reversalflowfield}
\end{figure}

The flow field during a typical reversal are drawn in Fig.~\ref{fig:reversalflowfield}.
The LSC accompanied with two small rolls near the diagonally opposing corners
exists at most times
in the same manner as shown in Fig.~\ref{fig:reversalflowfield}(a),
or its reflection with respect to the vertical axis $x=0$.
In Fig.~\ref{fig:reversalflowfield}(a),
the LSC is identified as the clockwise roll including the origin,
and the two corner rolls are as the counter-clockwise ones near the bottom right corner and the top left corner in the square domain.
The velocity field demonstrates that
the LSC elevates the hot fluid to the top
and drops the cold fluid to the bottom
by helping the growth of the plumes along the sidewalls from the bottom left and the top right,
respectively.
Therefore, the LSC plays a major role in heat transfer between the top and bottom.
On the other hand,
the corner rolls have little contribution to the heat transfer
because each corner roll does not touch both top and bottom boundaries.

Slightly before the reversal $t=490$,
the two corner rolls,
especially the boundary flows near the sidewalls evolve,
and the two corner rolls make the LSC small (Fig.~\ref{fig:reversalflowfield}(b)).
The two corner rolls further evolve,
and pinch the LSC (Fig.~\ref{fig:reversalflowfield}(c)).
At this moment $t=506$,
$L$ changes its sign to negative owing to the strong counter-clockwise corner rolls.
At $t=509$,
the two corner rolls touch with each other near the origin,
and tear the LSC into two rolls.
Subsequently, these rolls exhibit the quadruple rolls (Fig.~\ref{fig:reversalflowfield}(d)).
At $t=512$,
the hot corner roll that was originally near the bottom right corner
has a strong vertical boundary flow along the right wall because of the buoyancy.
The vertical boundary flow is separated from the sidewall
by the clockwise roll near the top right corner
that was a part of the clockwise LSC,
and the separation takes the hot roll to the region near the origin.
The hot roll lies on the cold roll that is similarly transformed from the cold corner roll that was originally near the top left corner (Fig.~\ref{fig:reversalflowfield}(e)).
They merge into a new counter-clockwise LSC near the origin.
As the new LSC evolve, the two rolls made of the former clockwise LSC shrink (Figs.~\ref{fig:reversalflowfield}(f)--(g)),
and settle to the top right and bottom left corners (Fig.~\ref{fig:reversalflowfield}(h)).
(See also a supplementary movie~\cite{supplementalmovie}.)

\begin{figure}
\includegraphics[width=.95\textwidth]{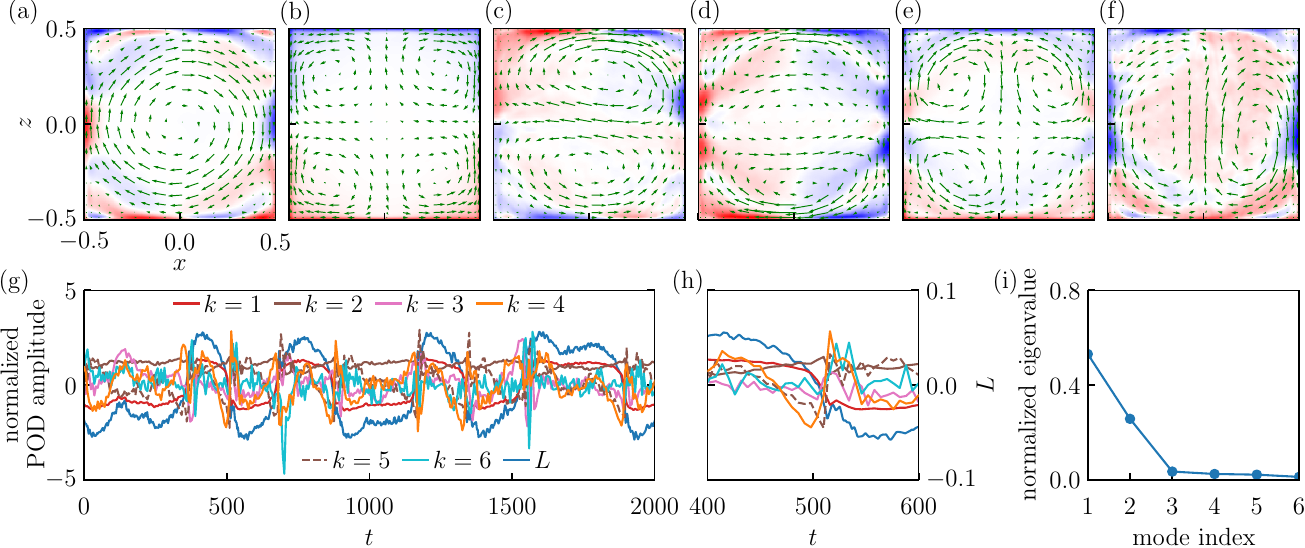}
\caption{
POD analysis.
(a)--(f) First six energetic POD modes.
(g) Normalized POD amplitudes and (h) those near a reversal.
(i) Normalized POD eigenspectrum.
}
\label{fig:pod}
\end{figure}
To characterize the spatio-temporal structures of the intermittent reversals,
the velocity and temperature fields are analyzed by using snapshot POD~\cite{sirovich1987turbulence,doi:10.2514/1.J056060}.
The snapshot POD finds
the eigenvalues $\lambda_k$
and the POD eigenmodes $\bm{\phi}_k = V \bm{\psi}_k / \sqrt{\lambda_k}$
by solving the eigenvalue problem $V^{\mathrm{T}} W_{\mathrm{GL}} V \bm{\psi}_k = \lambda_k \bm{\psi}_k$.
Here,
the matrix $V \in \mathbb{R}^{3N_{\mathrm{GL}} \times N_{\mathrm{ss}}}$ consists of the state variable $\bm{v}=(\bm{u},\theta)$,
and the weight matrix
$W_{\mathrm{GL}} \in \mathbb{R}^{3N_{\mathrm{GL}} \times 3N_{\mathrm{GL}}}$
is introduced
so that the weighted squared sum of each column of $V$
defined on the nonuniform Gauss-Lobatto points
is proportional to sum of the kinetic energy and the temperature variance,
which is referred to as total ``POD energy''.
The number of the snapshots is $N_{\mathrm{ss}}=2000$,
and the snapshots are taken at every $6$ time units in the numerical simulation.

The first six energetic modes are drawn in Fig.~\ref{fig:pod}(a)--(f).
The first mode has a single roll with thermal flux along the sidewalls,
accounting mainly for the LSC
(mode L; Fig.~\ref{fig:pod}(a)).
The second mode has quadruple rolls with mean temperature field
(mode Q; Fig.~\ref{fig:pod}(b)).
The sum of the first and second modes makes
the secondary rolls near diagonally opposing two corners,
and it also makes the LSC large near the other diagonally opposing two corners.
The clockwise LSC and the counter-clockwise corner rolls near the bottom right corner and the top left corner
similar to Fig.~\ref{fig:pod}(a)
are produced by $a_1 \bm{\phi}_1 + a_2 \bm{\phi}_2$,
where the amplitudes are supposed to be $a_1 > a_2 > 0$.
Their reflection with respect to the vertical axis $x=0$,
i.e., the counter-clockwise LSC and the clockwise corner rolls near the bottom left corner and the top right corner
is produced by $-a_1 \bm{\phi}_1 + a_2 \bm{\phi}_2$.
The third mode has two horizontal rolls
(mode S; Fig.~\ref{fig:pod}(c)).
The fourth mode has two co-rotating rolls and a small counter-rotating roll
with oppositely signed strong fluxes near the sidewalls
(mode L$_{\ast}$; Fig.~\ref{fig:pod}(d)).
The fifth mode is another quadruple-roll mode
(mode Q$_{\ast}$; Fig.~\ref{fig:pod}(e)).
The sixth mode has two vertical rolls
(mode S$_{\ast}$; Fig.~\ref{fig:pod}(f)).
The first six modes of the POD capture more than 88\% of the total POD energy.
The mode names except the mode Q$_{\ast}$ followed those in Ref.~\cite{PhysRevE.95.013112}.

The time series of the normalized POD amplitudes of the first six energetic modes are drawn in Figs.~\ref{fig:pod}(g) and (h).
Obviously,
the amplitude of the mode L ($k=1$) is synchronous with $L$.
The slow increase of the amplitude of the mode Q ($k=2$) in the quasi-stable state
comes from the growth of the secondary rolls as well as the erosion of the LSC.
The amplitude of the mode Q rapidly decreases just after the reversal.
Similarly,
the mode L$_{\ast}$ ($k=4$) slowly changes its amplitude with sign inversion
in the quasi-stable state,
and the amplitude rapidly inverts its sign during the reversal
followed by another relatively slow inversion after the reversal.

These POD analysis agrees with that shown in Ref.~\cite{PhysRevE.95.013112}
except that the mode S$_{\ast}$ appears as the fifth mode in Ref.~\cite{PhysRevE.95.013112}.
This alteration of the order results from the smallness of the two eigenvalues.
In fact,
the eigenspectrum shown in Fig.~\ref{fig:pod}(i) illustrates
that the eigenvalues of the first two modes are outstandingly large,
and those of the rest are small.

The time variations of the modes Q and L$_{\ast}$ are large
near the times when the angular momentum $L$ inverts its sign,
and the modes Q and L$_{\ast}$ have characteristic structures near the sidewalls.
It indicates that
the angular momentum $L$ is predictable
if the variations of the flow near the sidewalls are measured at each time.

\begin{figure}
\includegraphics[width=.9\textwidth]{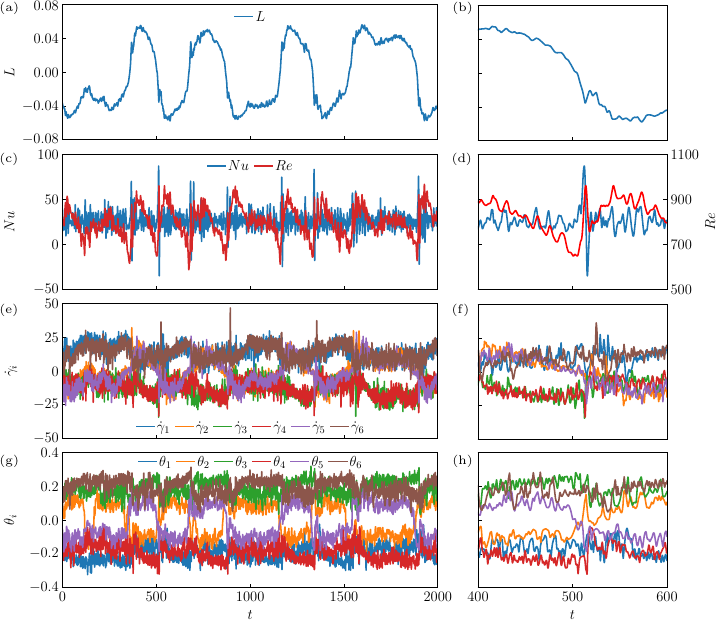}
\caption{
Time evolution of
(a) total angular momentum $L$,
(c) Nusselt number $Nu$ and Reynolds number $Re$,
(e) shear rates at six locations on the sidewalls $\dot{\gamma}_i$,
and
(g) temperatures at the six locations $\theta_i$.
(b, d, f, h) Time evolution near a reversal.
The linear weighted moving average is applied to smooth the values.
}
\label{fig:timeseries}
\end{figure}

The time series of the total angular momentum $L$ in Figs.~\ref{fig:timeseries}(a)--(b)
displays that
$L$ does not change its sign over long intervals,
and intermittently changes its sign in short times.
In the quasi-stable states,
$|L| \approx 0.04$ is almost constant.
The typical inter-reversal time,
which is the typical duration of the quasi-stable state,
is roughly evaluated as $200$,
and the sign inversions of $L$ are completed in short times of about $30$
(See also Figs.~\ref{fig:reversalflowfield}(b)--(g)).
During the early part in the quasi-stable state,
$|L|$ slowly decreases
as the temperature distribution in the new LSC matches that outside of the LSC,
and the secondary rolls are formed (See also Figs.~\ref{fig:reversalflowfield}(f)--(h)).
During the late part,
$|L|$ slowly decreases as the growing secondary rolls penetrate the LSC.
The growth is not necessarily monotonic,
and the secondary rolls sometimes decay,
resulting in a longer interval of sign invariance of $L$.

The turbulent RBC is characterized by two non-dimensional numbers:
the Nusselt number $Nu$ and the Reynolds number $Re$.
These non-dimensional numbers are defined as
\begin{eqnarray}
 Nu = \sqrt{Ra Pr} \langle u_z \theta \rangle_V + 1
 ,
\qquad
 Re = \sqrt{\frac{Ra}{Pr}} \sqrt{\langle |\bm{u}|^2 \rangle_V}
 ,
\end{eqnarray}
and their time series are drawn in Figs.~\ref{fig:timeseries}(c)--(d).
Here, $\langle \cdot \rangle_V$ represents averaging over the numerical domain.
When the LSC exists, $Nu$ fluctuates around $25$,
and $Re$ decreases.
Nearly the time of the sign change of $L$,
local minimal $Re$,
which is proportional to the square root of the total kinetic energy,
appears (Fig.~\ref{fig:reversalflowfield}(c)).
A positive spike of $Nu$ at $t=512$ (Fig.~\ref{fig:reversalflowfield}(e)),
a spike of $Re$ at $t=514$ (Fig.~\ref{fig:reversalflowfield}(f)),
and a negative spike of $Nu$ at $t=516$ (Fig.~\ref{fig:reversalflowfield}(g))
occur in short succession.
The hot roll that was the counter-clockwise roll near the bottom right corner
and the cold roll that was near the top left corner
merge into the new LSC at $t=512$.
At this moment,
the strong boundary flows elevate the hot fluid and drop the cold fluid,
and the positive correlation between $u_z$ and $\theta$ in $Nu$
is large to make the positive spike of $Nu$.
The buoyancy accelerates the boundary flows and hence the LSC,
resulting in the increase of the kinetic energy and hence $Re$ at $t=514$.
The horizontal flows move the hot fluid in the newly-formed LSC to the left
and the cold fluid to the right at $t=516$.
At this moment,
the LSC drops the hot fluid and elevates the cold fluid,
and the negative correlation between $u_z$ and $\theta$ in $Nu$
makes the negative spike of $Nu$,
and even $Nu<0$ sometimes appears.
The local Nusselt numbers averaged over the horizontal direction
near the top and the bottom are positive
even when the global $Nu$ are instantaneously negative~\cite{PhysRevLett.110.114503},
and the negative global $Nu$ at instantaneous times
is caused transiently by the reversal of the LSC,
and does not contradict the second law of thermodynamics.

The time series of the shear rates and the temperatures
at the six locations on the sidewalls
are respectively drawn in Figs.~\ref{fig:timeseries}(e)--(f) and Figs.~\ref{fig:timeseries}(g)--(h).
The time series of $\dot{\gamma}_i$ and $\theta_i$ respectively show
$\dot{\gamma}_1 \approx \dot{\gamma}_6$, $\dot{\gamma}_2 \approx \dot{\gamma}_5$, and $\dot{\gamma}_3 \approx \dot{\gamma}_4$,
and
$\theta_1 \approx -\theta_6$, $\theta_2 \approx -\theta_5$, and $\theta_3 \approx -\theta_4$,
though the fluctuations are large.
These symmetries of the time series are derived from the centrosymmetry,
which results in the flow field invariant under the rotation about the origin
\begin{eqnarray}
 R_{\pi}:
 \begin{bmatrix}
  u_x \\
  u_z \\
 \theta
 \end{bmatrix}
 (x,z)
\longrightarrow
 \begin{bmatrix}
  -u_x \\
  -u_z \\
  -\theta
 \end{bmatrix}
 (-x,-z)
.
\end{eqnarray}
Because the first two four energetic modes, L and Q,
as well as two other modes, L$_{\ast}$ and Q$_{\ast}$,
have the centrosymmetry,
the centrosymmetry is almost satisfied by the LSC and the two corner rolls.
Such centrosymmetry is seen in the flow fields in Fig.~\ref{fig:reversalflowfield}
and a supplementary movie~\cite{supplementalmovie}.

The reversal can be seen in the time series of the shear rates and the temperatures
on the sidewalls.
Let us examine the reversal from a clockwise LSC ($L>0$)
to a counter-clockwise one ($L<0$) near $t=500$ shown in Figs.~\ref{fig:timeseries}(f) and (h)
for an example.
The clockwise LSC is accompanied with
the hot counter-clockwise roll near the bottom right corner
and the cold one near the top left corner before the reversal.
Because the clockwise LSC touches $z=0$ on both sidewalls,
$\dot{\gamma}_2$, $\dot{\gamma}_5$, $-\theta_2$, and $\theta_5$
are mostly synchronous with $L$.
The shear rates and the temperatures at locations 2 and 5
are respectively close to those at 1 and 6
because the LSC touches the walls at the locations 1, 2, 5 and 6 at this time.
(See also Fig.~\ref{fig:reversalflowfield}(a).)
The two corner rolls and the boundary layers on the sidewalls
slowly develop.
Then,
$\dot{\gamma}_3$ and $\dot{\gamma}_4$ increase ($|\dot{\gamma}_3|$ and $|\dot{\gamma}_4|$ decrease and $\dot{\gamma}_3, \dot{\gamma}_4<0$),
while $\theta_3$ increases and $\theta_4$ decreases.
As the corner rolls grow further,
they cross the horizontal axis $z=0$.
Then,
the shear rates and the temperatures at locations 2 and 5
become closer to those at 3 and 4, respectively.
(See also Fig.~\ref{fig:reversalflowfield}(d).)
During the opposite reversal from a counter-clockwise LSC to a clockwise one,
similar evolution can be seen
in the shear rates and the temperatures at the locations 1, 2, 5 and 6.

\begin{figure}
\includegraphics[width=.9\textwidth]{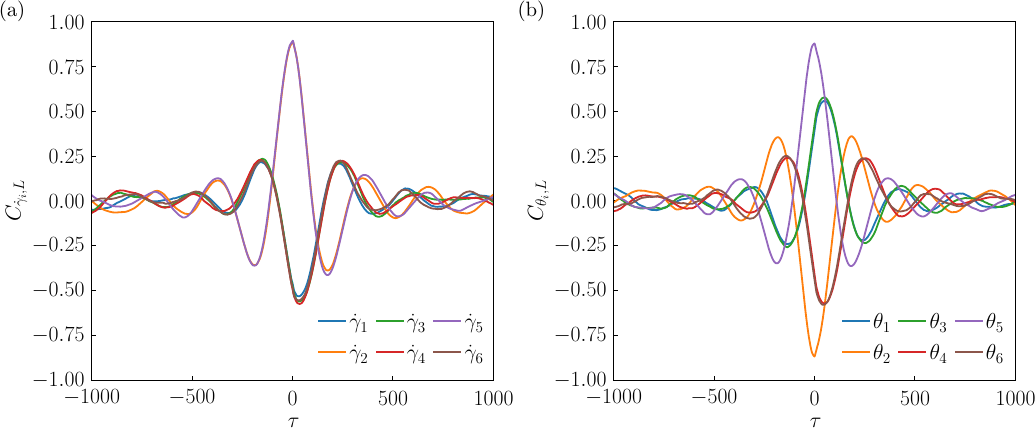}
\caption{Cross-correlations 
 (a) between $\dot{\gamma}_i$ and $L$,
 and (b) between $\theta_i$ and $L$.
}
\label{fig:crosscorrelation}
\end{figure}
The cross-correlations between $\dot{\gamma}_i$ and $L$,
and between $\theta_i$ and $L$ are defined as
\begin{eqnarray}
 C_{\dot{\gamma}_i, L}(\tau) = \frac{\int (\dot{\gamma}_i(t+\tau) -\overline{\dot{\gamma}}_i) (L(t) - \overline{L}) dt}%
 {\sqrt{\int (\dot{\gamma}_i(t) -\overline{\dot{\gamma}}_i)^2 dt \int (L(t) - \overline{L})^2 dt}}
 ,
\qquad
 C_{\theta_i, L}(\tau) = \frac{\int (\theta_i(t+\tau) -\overline{\theta}_i) (L(t) - \overline{L}) dt}%
 {\sqrt{\int (\theta_i(t) -\overline{\theta}_i)^2 dt \int (L(t) - \overline{L})^2 dt}}
,
\end{eqnarray}
where $\overline{\  \cdot \ }$ represents time averaging.
The strong synchronization of $\dot{\gamma}_2$, $\dot{\gamma}_5$, $-\theta_2$, and $\theta_5$ with $L$
is confirmed in Fig.~\ref{fig:crosscorrelation},
which illustrates that the cross-correlations are large or negatively large at $\tau \approx 0$.
In addition,
the cross-correlations also demonstrate that
the time delays of the shear rates and the temperatures at the locations 1, 3, 4, and 6 with respect to $L$
are roughly evaluated as $\tau \approx 60$
because the secondary rolls near the diagonally opposing corners are
basically driven by the LSC except the onset of the reversals.
Therefore,
time series of the shear rates and the temperatures on the sidewalls,
especially those at the center of the sidewalls ($z=0$)
reflect the LSCs in the bulk region.
In fact, the reversals of LSCs in a three-dimensional flow had been detected
mainly by measurements of the temperature at several points on the sidewalls~\cite{brown_ahlers_2006}.

\begin{figure}
\includegraphics[width=.9\textwidth]{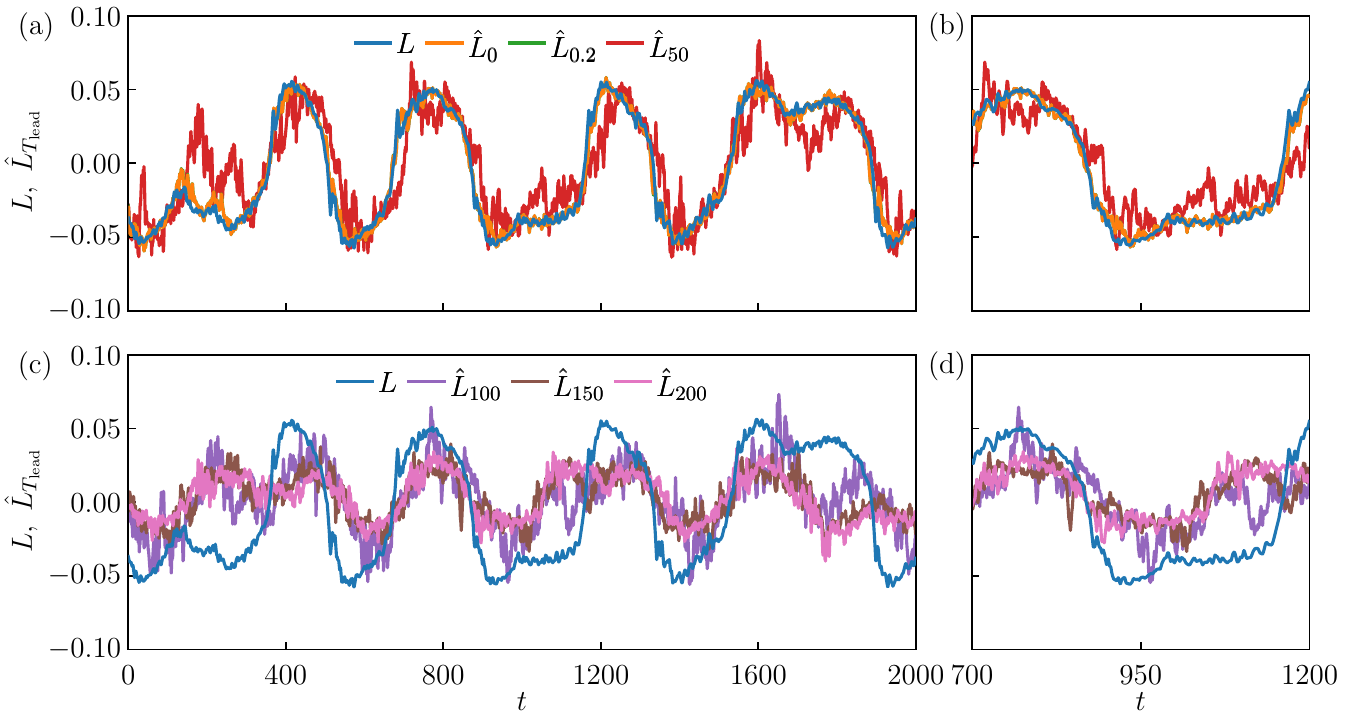}
\caption{
Time evolution of $L$ and its prediction with the lead times
(a) $T_{\mathrm{lead}}=0$, $0.2$, $50$,
and (c) $T_{\mathrm{lead}}=100$, $150$, $200$.
(b, d) Enlargement in the time window $[700, 1200]$
that has a typical reversal followed by a reversal having a long inter-reversal time.
}
\label{fig:prediction}
\end{figure}

In this paper,
the shear rates and temperatures at the six locations on the sidewalls
are employed as input for the RC prediction of the reversals.
The time series of $L$ and its prediction with the several lead times, $\hat{L}_{T_{\mathrm{lead}}}$,
are drawn in Fig.~\ref{fig:prediction}.
As written in Sec.~\ref{sec:numericalmethod},
$\hat{L}_{T_{\mathrm{lead}}}(t)$ is predicted from
$\dot{\gamma}_i(t-T_{\mathrm{lead}})$ and $\theta_i(t-T_{\mathrm{lead}})$
with lead time $T_{\mathrm{lead}}$.

The simultaneous and single-step-ahead predictions,
$\hat{L}_{0}$ and $\hat{L}_{0.2}$,
are almost indistinguishable from each other,
and both predictions mostly follow $L$
with the fluctuation resulting from the large fluctuation of the input vector,
i.e., $\dot{\gamma}_i$ and $\theta_i$.
The short-term prediction $\hat{L}_{50}$ is often delayed from $L$,
and the delay is noticeable near the reversals.
As the lead time $T_{\mathrm{lead}}$ increases,
the difference between $\hat{L}_{T_{\mathrm{lead}}}$ and $L$ becomes large.
In particular,
in the late part of the long quasi-stable state,
$\hat{L}_{T_{\mathrm{lead}}}$ with $T_{\mathrm{lead}}\geq 100$ has the sign opposite to $L$
predicting false reversals.
Therefore,
the prediction $\hat{L}_{T_{\mathrm{lead}}}$ with long lead times $T_{\mathrm{lead}}$
shows an almost periodic oscillation with the period roughly evaluated as $200$,
which is close to the mean inter-reversal time.
After the long quasi-stable state,
$\hat{L}_{T_{\mathrm{lead}}}$ with large $T_{\mathrm{lead}}$ again comes close to $L$
without any feedback mechanism.
It is due to the echo state property stating that
the reservoir dynamics should asymptotically depend only on the driving input.
In addition,
the amplitudes of the large-scale oscillation of $\hat{L}_{T_{\mathrm{lead}}}$ decrease,
and $\hat{L}_{T_{\mathrm{lead}}}$ is flattened,
as $T_{\mathrm{lead}}$ increases.
The cross-correlation between $L(t+T_{\mathrm{lead}})$ and $\bm{r}(t)$ decreases
as $T_{\mathrm{lead}}$ increases,
because $\bm{r}(t)$ depends on $\dot{\gamma}_i(t)$ and $\theta_i(t)$.
According to Eq.~(\ref{eq:Woutopt}),
the small cross-correlation between $L(t+T_{\mathrm{lead}})$ and $\bm{r}(t)$,
i.e., small $DR^{\mathrm{T}}$
makes each element of $W_{\mathrm{out}}^{\mathrm{opt}}$ small.
Note that $(R R^{\mathrm{T}} + \beta I)^{-1}$ is independent of $T_{\mathrm{lead}}$.
It results in the small amplitude of the prediction.

Figures~\ref{fig:prediction}(b) and (d) show the time series of $L$ and its prediction
in a typical-duration quasi-stable state in $t < 850$
and
a long quasi-stable state in $850 < t < 1150$.
One can clearly observe in this enlargement that
the signs of $\hat{L}_{T_{\mathrm{lead}}}$ with $T_{\mathrm{lead}}\geq 100$
agree with that of $L$
in the typical-duration quasi-stable state
and until the mean inter-reversal time after a reversal in the long quasi-stable state $850 < t < 1050$.
As written above,
after the mean inter-reversal time in the long quasi-stable state $t > 1050$,
$\hat{L}_{T_{\mathrm{lead}}}$ with $T_{\mathrm{lead}}\geq 100$ and $L$
have the opposite signs.

\begin{figure}
\includegraphics[width=.9\textwidth]{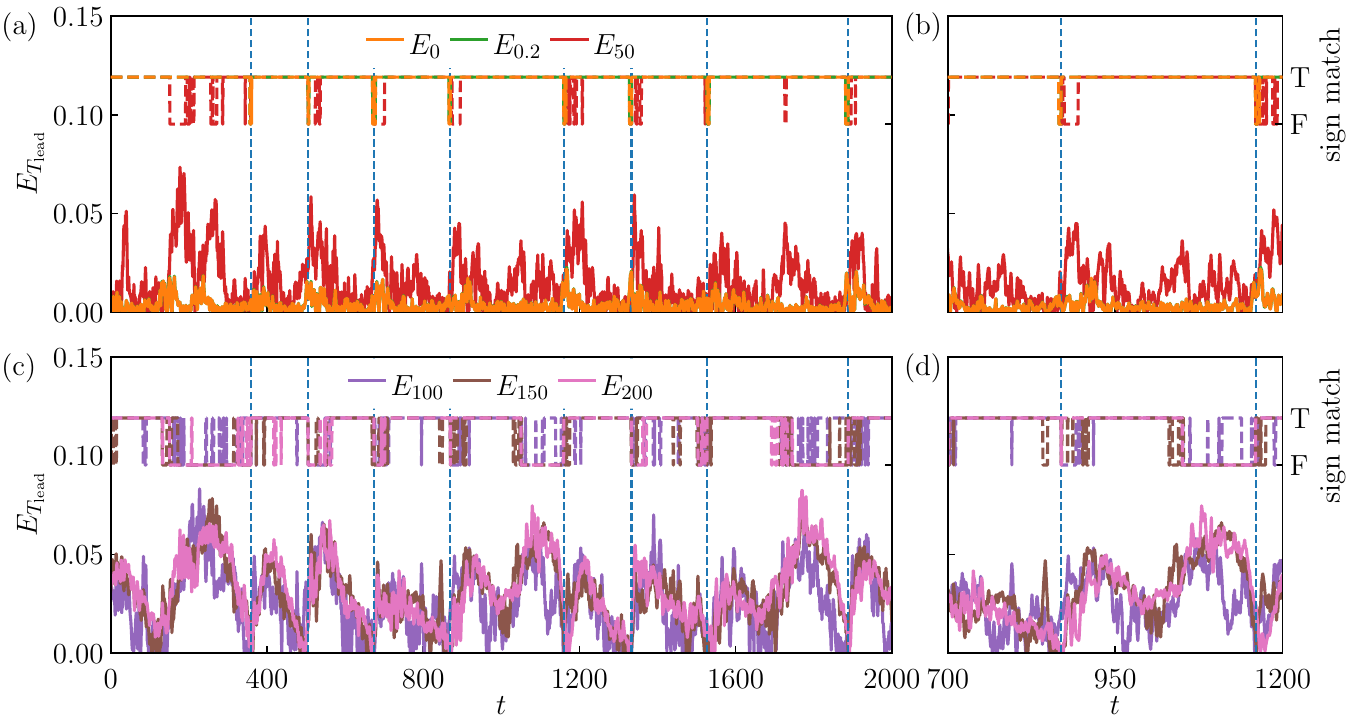}
\caption{
Prediction error and sign match
between $\hat{L}_{T_{\mathrm{lead}}}$ and $L$.
(a) $T_{\mathrm{lead}}=0$, $0.2$, $50$,
and (c) $T_{\mathrm{lead}}=100$, $150$, $200$.
(b, d) Enlargement in the time window $[700, 1200]$.
The letters T and F respectively represent that
the signs of $\hat{L}_{T_{\mathrm{lead}}}$ and $L$ match and do not match.
The blue dashed lines represent the times of $L=0$.
}
\label{fig:errorseries}
\end{figure}

\begin{figure}
\includegraphics[width=.9\textwidth]{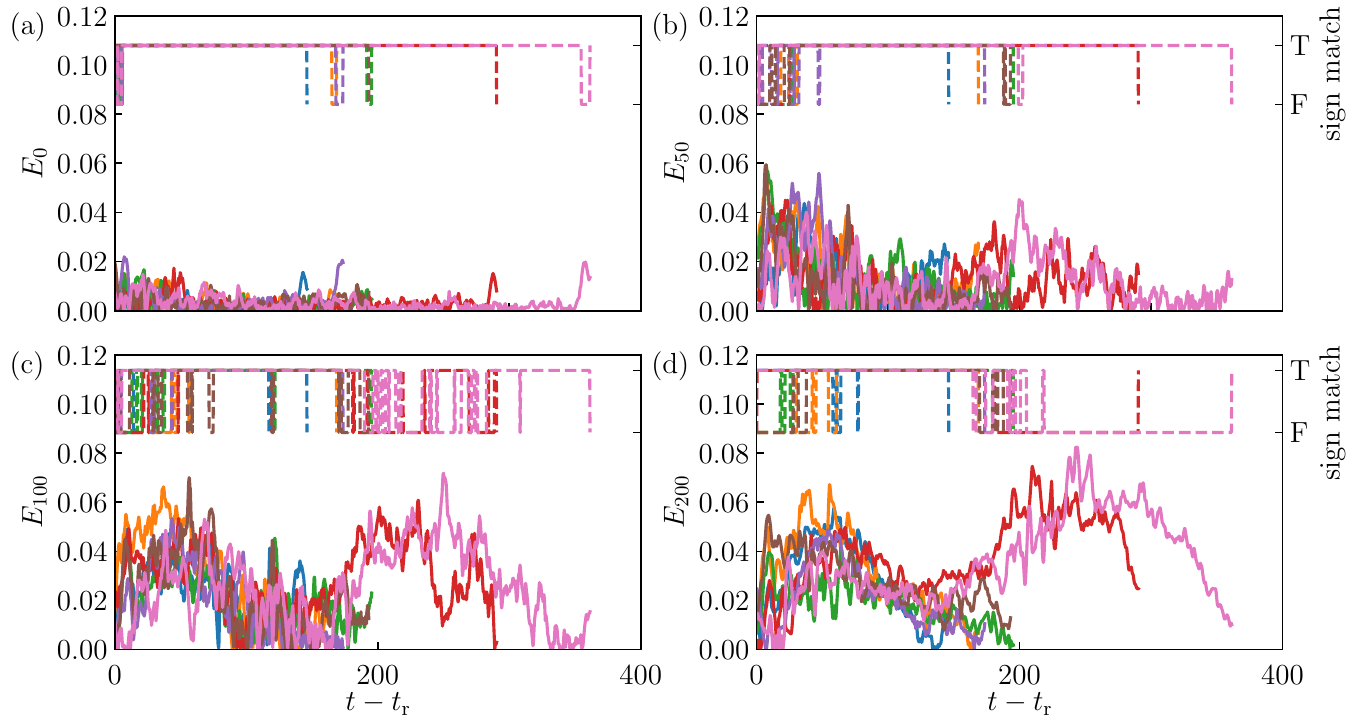}
\caption{
Time series of the prediction error and sign match between $\hat{L}_{T_{\mathrm{lead}}}$ and $L$ after the reversals.
(a) $T_{\mathrm{lead}}=0$, (b) $50$, (c) $100$, and (d) $200$.
The abscissa $t-t_{\mathrm{r}}$ represents the time
elapsed after the reversals occurring at $t_{\mathrm{r}}$.
The first seven reversals denoted in Fig.~\ref{fig:errorseries} are designated
as the times of the origin in this figure.
See also the caption to Fig.~\ref{fig:errorseries}.
}
\label{fig:errorafterreversal}
\end{figure}

To see the prediction error more clearly,
the time series of the error $E_{T_{\mathrm{lead}}} = |\hat{L}_{T_{\mathrm{lead}}} - L|$
and the sign matches between $\hat{L}_{T_{\mathrm{lead}}}$ and $L$
are drawn in Fig.~\ref{fig:errorseries},
and those with the time elapsed after the reversals until the subsequent reversals
are also drawn in Fig.~\ref{fig:errorafterreversal}.
Here,
the letters T and F in Figs.~\ref{fig:errorseries} and \ref{fig:errorafterreversal}
respectively represent that
the signs of $\hat{L}_{T_{\mathrm{lead}}}$ and $L$ match and do not match,
and $t_{\mathrm{r}}$ denotes the times of the reversals.

The prediction errors of the simultaneous and single-step-ahead predictions,
$T_{\mathrm{lead}} = 0$ and $0.2$,
are almost indistinguishable similar to Fig.~\ref{fig:prediction},
and the errors are small.
The relatively large errors and the sign mismatches
come from the rapid change of $L$ during the reversals.
The errors of $\hat{L}_{50}$ are mostly small,
but the large errors just after the reversals are caused by the delay of $\hat{L}_{50}$ from $L$.
Moreover,
the large errors emerge also in the late part of the long quasi-stable states.
The errors $\hat{L}_{T_{\mathrm{lead}}}$ with $T_{\mathrm{lead}} \geq 100$
are large
when $t-t_{\mathrm{r}} \lessapprox 50$ and $t-t_{\mathrm{r}} \gtrapprox 200$.
The former comes from the small amplitudes of $\hat{L}_{T_{\mathrm{lead}}}$,
and the latter is due to false reversals in the late part of the long quasi-stable states.
On the other hand,
these errors are small in $50 \lessapprox t-t_{\mathrm{r}} \lessapprox 200$.
It must be noted here that
such errors with $T_{\mathrm{lead}}=200$ in $50 \lessapprox t-t_{\mathrm{r}} \lessapprox 200$ are small,
though $\hat{L}_{200}$ is predicted from the inputs at a time in the previous quasi-stable state.

\begin{figure}
\includegraphics{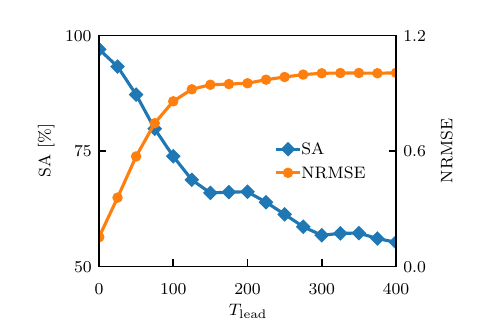}
\caption{
Sign accuracy (SA)
and normalized root-mean-square error (NRMSE)
between $\hat{L}_{T_{\mathrm{lead}}}$ of $L$
against $T_{\mathrm{lead}}$.
}
\label{fig:sanrmse}
\end{figure}

Figure~\ref{fig:sanrmse} shows
the dependence of the sign accuracy (SA) of $L$ and $\hat{L}_{T_{\mathrm{lead}}}$
as well as the normalized root-mean-square error (NRMSE)
on the lead times.
Here, the SA and the NRMSE are respectively defined as
\begin{eqnarray}
A_{\mathrm{sign}}(\hat{L}_{T_{\mathrm{lead}}}; L) = \left(\frac{1}{T_{\mathrm{test}}} \int_0^{T_{\mathrm{test}}} H(\hat{L}_{T_{\mathrm{lead}}}(t) L(t)) dt\right) \times 100[\mathrm{\%}]
,
\,
E(\hat{L}_{T_{\mathrm{lead}}}; L) = \left(\frac{\int_0^{T_{\mathrm{test}}} (\hat{L}_{T_{\mathrm{lead}}}(t) - L(t))^2 dt}{\int_0^{T_{\mathrm{test}}} (L(t) - \overline{L})^2 dt}\right)^{1/2}
,
\end{eqnarray}
where $H$ denotes the Heaviside step function.
The SAs of the short-term prediction where $T_{\mathrm{lead}} \leq 0.2$ are approximately $97$\%,
and the simultaneous and single-step-ahead predictions successfully reproduce the reversals of the LSC.
The $3$\% difference of the signs comes from the fluctuation of the prediction values near the reversals $L=0$.
The large fluctuation of the prediction values also makes the NRMSE as large as $0.15$.
The SA displays that
$\hat{L}_{T_{\mathrm{lead}}}$ with $T_{\mathrm{lead}}\leq 100$ have
the same signs as $L$ has
more than $74$\%.
These differences between $\hat{L}_{T_{\mathrm{lead}}}$ and $L$ make the decrease of the SA
and the increase of the NRMSE with the increase of $T_{\mathrm{lead}}$.
As $T_{\mathrm{lead}}$ further increases,
the SA gradually decreases toward 50\%, which corresponds to the coin toss.
The plateaus appear every $180$ time units,
which is close to the mean inter-reversal time and the period of $\hat{L}_{T_{\mathrm{lead}}}$.
The NRMSE reaches at unity
for $T_{\mathrm{lead}}$ smaller than the SA reaches at its saturation.
The prediction of the value of $L$ is much more difficult than that of the reversals, i.e., the signs of $L$.
The prediction is statistically validated
by using the random permutation (shuffle and split) cross validation in Appendix~\ref{sec:crossvalidation}.

\section{Discussion}
\label{sec:discussion}

\begin{table}[b]
\caption{
Selection of inputs by sparse measurements.
\label{tab:sparsitycheck}
}
\begin{ruledtabular}
\begin{tabular}{lcc}
label & input elements & locations
\\
\colrule
14$\dot{\gamma}\theta$ & $\dot{\gamma}$, $\theta$ & $i=1, \ldots, 6$, and $(x,z)=(\pm 1/2, \pm\sqrt{2 \pm \sqrt{2}}/2)$
\\
6$\dot{\gamma}\theta$ & $\dot{\gamma}$, $\theta$ & $i=1, \ldots, 6$
\\
6$\dot{\gamma}$ & $\dot{\gamma}$ & $i=1, \ldots, 6$
\\
6$\theta$ & $\theta$ & $i=1, \ldots, 6$
\\
4$\dot{\gamma}\theta$ & $\dot{\gamma}$, $\theta$ & $i=1,3,4,6$
\\
3$\dot{\gamma}\theta$ & $\dot{\gamma}$, $\theta$ & $i=1,2,3$
\\
2$\dot{\gamma}\theta$ & $\dot{\gamma}$, $\theta$ & $i=2, 5$
\\
\end{tabular}
\end{ruledtabular}
\end{table}

\begin{figure}
 \centering
 \includegraphics[width=.85\textwidth]{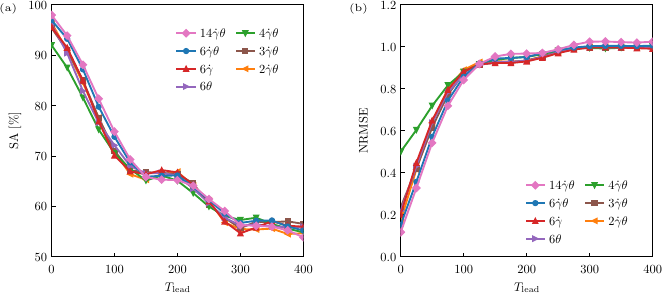}
\caption{
Dependence of (a) SA and (b) NRMSE
 of $L$ and $\hat{L}_{T_{\mathrm{lead}}}$
on selection of the elements of the input vector.
See Table~\ref{tab:sparsitycheck} for the legends.
}
\label{fig:sparsitycheck}
\end{figure}

It is expected that
the more elements of the input produce the better prediction,
but the accuracy should saturate if sufficient elements are provided.
Figure~\ref{fig:sparsitycheck} demonstrates
the SA and the NRMSE for some selections of the inputs,
which are summarized in Table~\ref{tab:sparsitycheck},
without changing the hyper-parameters.
Note that the input used in the main results is labeled as $6\dot{\gamma}\theta$.
Even if eight locations are added ($14\dot{\gamma}\theta$),
the increase of the accuracy from that with the input used in the main results ($6\dot{\gamma}\theta$)
is not large.
If the locations only at the center of the sidewalls ($2\dot{\gamma}\theta$)
or those only on one sidewall ($3\dot{\gamma}\theta$) are used,
the accuracy decreases from that with the six locations ($6\dot{\gamma}\theta$).
Similarly,
if only the shear rates ($6\dot{\gamma}$) or the temperatures ($6\theta$) are used in the input,
the accuracy decreases.
Moreover,
if the locations at the center of the sidewalls are excluded ($4\dot{\gamma}\theta$),
the accuracy largely decreases.
It is consistent with
the strong synchronization of $\dot{\gamma}_2$, $\dot{\gamma}_5$, $-\theta_2$, and $\theta_5$ with $L$ observed in Figs.~\ref{fig:timeseries} and \ref{fig:crosscorrelation}.
Therefore,
the input consisting of the shear rates and temperatures at the six locations ($6\dot{\gamma}\theta$)
are effective for the RC prediction.

As seen in Fig.~\ref{fig:prediction},
the short-term prediction successfully reproduces the large-scale oscillation of $L$,
and the reversals having a typical inter-reversal time can be successfully predicted
even by the long-term prediction.
The success of the prediction by using the sparse measurable input
reveals that similar prediction is feasible in laboratory experiments.
On the other hand,
the angular momentum generated by the long-term prediction
deviates from $L$ in the late part of the long quasi-stable state.

\begin{figure}
\includegraphics[width=.9\textwidth]{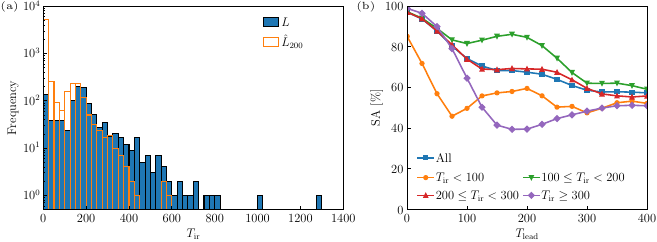}
\caption{
(a) Histograms of lengths of the inter-reversal time $T_{\mathrm{ir}}$
of $L$ and $\hat{L}_{200}$
 in a time interval $2 \times 10^5$.
(b) Conditional SA of $L$ and $\hat{L}_{T_{\mathrm{lead}}}$
for $T_{\mathrm{ir}} < 100$,
 $100 \leq T_{\mathrm{ir}} < 200$,
 $200 \leq T_{\mathrm{ir}} < 300$,
 and $T_{\mathrm{ir}} \geq 300$.
Each SA is calculated in 100 inter-reversal times.
}
\label{fig:conditional}
\end{figure}

To see the possibility of the long-term prediction,
the dependence of the predictability on the length of the inter-reversal times $T_{\mathrm{ir}}$ is examined.
The lengths of the inter-reversal times are widely distributed
as shown in their histogram (Fig.~\ref{fig:conditional}(a)).
Short inter-reversal times $T_{\mathrm{ir}} < 25$ come mainly from the fluctuation near the reversals $L=0$.
The statistical mode of $T_{\mathrm{ir}}$ appears near $200$,
indicating that the reversals occur as frequently as once in $200$ time units.
The exponential decay at the tail in the histogram suggests that
the reversals follow a Poisson statistics,
which agrees with the experimental measurements in a cylindrical container~\cite{PhysRevLett.95.084503,10.1063/1.2920444}.
The Poisson distribution indicates that
the reversals from one state to another in 2D RBC are independent from each other,
and it is confirmed by the embedding method~\cite{10.1063/1.5081031}.
The independence implies that
the prediction in one quasi-stable state is possible,
but that after the reversal is not.
The long-term prediction mostly provides
$\hat{L}_{T_{\mathrm{lead}}}$ whose sign matches $L$
in the early part of the long quasi-stable state,
but it simply results from the average behavior with the period equal to the typical-duration quasi-stable state.
In fact,
the histogram of $\hat{L}_{200}$ is more short-tailed than that of $L$.
In this sense,
the long-term prediction overfits the reversals having the typical inter-reversal time.
In addition,
it also indicates that
the long-term prediction with a long lead time could be more successful for larger $Ra$'s
because the inter-reversal times increase
as $Ra$ increases~\cite{PhysRevLett.105.034503}.

The dependence of the predictability on $T_{\mathrm{ir}}$
is examined by the SA under an {\itshape a posteriori} condition
classifying the SA according to $T_{\mathrm{ir}}$.
The conditional SA of $L(t + T_{\mathrm{lead}})$ and $\hat{L}_{T_{\mathrm{lead}}}(t + T_{\mathrm{lead}})$
predicted from $\dot{\gamma}_i(t)$ and $\theta_i(t)$
where $t$ is within the quasi-stable states having
$T_{\mathrm{ir}} < 100$, $100 \leq T_{\mathrm{ir}} < 200$, $200 \leq T_{\mathrm{ir}} < 300$, and $T_{\mathrm{ir}} \geq 300$
are drawn in Fig.~\ref{fig:conditional}(b).
Each conditional SA is obtained from 100 inter-reversal times of $t$.

The conditional SA for $T_{\mathrm{ir}} < 100$ is smaller than those for $T_{\mathrm{ir}} \geq 100$ at small $T_{\mathrm{lead}} < 100$.
The difference comes mainly from
the fluctuation near the reversals $L=0$ in the short inter-reversal times $T_{\mathrm{ir}} < 25$.
The conditional SA for $T_{\mathrm{ir}} < 100$ goes to $50$\%,
which indicates the RC is trying to predict the outcome of the coin toss
due to the statistical independence of the flows after the reversals.
The relatively large SA for $T_{\mathrm{lead}} \leq 25$ indicates
the possibility of the prediction in the short quasi-stable states $25 < T_{\mathrm{ir}} < 100$,
which is not due to the fluctuation.

The conditional SAs for $T_{\mathrm{ir}} \geq 100$ almost collapse to that from all the inter-reversal times,
and these SAs are mostly larger than $80$\%
in $T_{\mathrm{lead}} \leq 75$.
Thus, the short-term prediction of the reversals is successful
independently of $T_{\mathrm{ir}}$ except the fluctuation near the reversals.

In the range $100 \leq T_{\mathrm{lead}} \leq 300$,
the conditional SAs for $T_{\mathrm{ir}} \geq 100$ clearly separate.
The conditional SA for $100 \leq T_{\mathrm{ir}} < 200$ starts to increase at $T_{\mathrm{lead}}=100$,
and has another peak, which is larger than 80\%,
at $T_{\mathrm{lead}} = 175$.
A similar but smaller peak appears also in the conditional SA for $200 \leq T_{\mathrm{ir}} < 300$
later at $T_{\mathrm{ir}} = 225$.
It results from the tendency
that the LSC are likely to rotate in the opposite direction after these peak times
because the statistical mode of the inter-reversal time is close to $200$.
On the other hand,
the conditional SA for $T_{\mathrm{ir}} \geq 300$ is smaller than 50\%,
and it is worse than the coin toss.
It results from the fact that
$\hat{L}_{T_{\mathrm{lead}}}$ has the opposite sign of $L$
in the late part of the long quasi-stable state (Figs.~\ref{fig:prediction}--\ref{fig:errorafterreversal}).
After $T_{\mathrm{lead}} > 300$,
all the conditional SAs converge to $50$\%,
which implies the coin toss,
after the repetition of the oscillation.

The two consecutive reversals are statistically independent from each other,
and the angular momentum in the subsequent quasi-stable state
is statistically independent of the flows in the preceding quasi-stable state.
The statistical independence is destructive,
and the reservoir computing cannot find any hints in the flows in the preceding
quasi-stable state.
Therefore,
the prediction after the reversal
tends to overfit the reversals having a typical inter-reversal time.
The SA for $T_{\mathrm{lead}} > 100$ is contaminated by the overfitting,
and the contamination is more remarkable in that for $T_{\mathrm{ir}} > 300$.
Such overfitting cannot be prevented by the Tikhonov regularization,
and the prediction accuracy is almost independent of the regularization parameter $\beta$
as shown in Appendix~\ref{sec:hyperparameters}.
The long-term prediction would be better
if we could restrict the training and the prediction only within the same quasi-stable states.

\section{Conclusion}
\label{sec:conclusion}

In this paper,
direct numerical simulations of intermittent reversals of LSCs in 2D turbulent RBC are performed.
It has been found from the DNS that
the growth of small counter-rotating secondary rolls near the diagonally opposing corners
triggers the reversals of the LSC,
and the boundary flows near the sidewalls
show characteristic structures near the time of the reversals.
The POD analysis confirmed these near-wall characteristic structures
of the velocity and the temperature.

LI-ESN,
which is a type of RC,
is used for the prediction of the total angular momentum
and hence the reversals of the LSC.
The shear rates and temperatures at the six locations on the sidewalls
are used as the elements of the sparse input of the locally measurable quantities
for the prediction.
The sparse input of the shear rates and temperatures on the sidewalls
can successfully reproduce the sign of the total angular momentum
with the precision higher than $80$\%
as long as the lead time is shorter than the half of the typical inter-reversal time, i.e., $T_{\mathrm{lead}} < 100$.
The shear rates on the sidewalls can be calculated from the wall shear stress,
and thus
both shear rates and temperatures are locally measurable by non-intrusive sensors installed in the sidewalls.
The successful prediction by using the non-intrusive sensing
provides the feasibility of the reservoir computing prediction in laboratory experiments and industrial applications,
and enables the closed-loop control of the LSC in RBC.
%
%
The long-term prediction is not successful
as it often provides the total angular momentum opposite in sign
in the late part of long quasi-stable states,
resulting from the contamination due to the statistical independence after the reversals during the training phase.
It is our future work to remove such contamination.

\begin{acknowledgments}
This work was partially supported by JSPS KAKENHI Grant No.~21K03883
and No.~22K03462.
This work was supported also by the Research Institute for Mathematical Sciences,
an International Joint Usage/Research Center located in Kyoto University.
M.I. wishes to acknowledge Mr.\ Yuto Enokido for insightful discussions.
\end{acknowledgments}

\appendix
\section{Hyper-parameter search of RC}
\label{sec:hyperparameters}

The hyper-parameters which potentially affect the prediction accuracy are
the number of reservoir nodes $N_{\mathrm{res}}$,
the spectral radius $\rho$ and the connectivity $C_{\mathrm{res}}$ of $W_{\mathrm{res}}$,
the leak ratio of LI-ESN $\alpha$,
the regularization parameter of the ridge regression $\beta$,
and the training length $T_{\mathrm{train}}$.
Because these are independent of each other,
the optimal values of these should be basically searched in the six-dimensional parameter space.
Instead of such time-consuming global search of the optimal values,
a sequential search of a small number of the hyper-parameters finds that
the hyper-parameter values used in the main text are nearly optimal ones
for $T_{\mathrm{lead}}=200$.

\begin{figure}
\includegraphics[width=.9\textwidth]{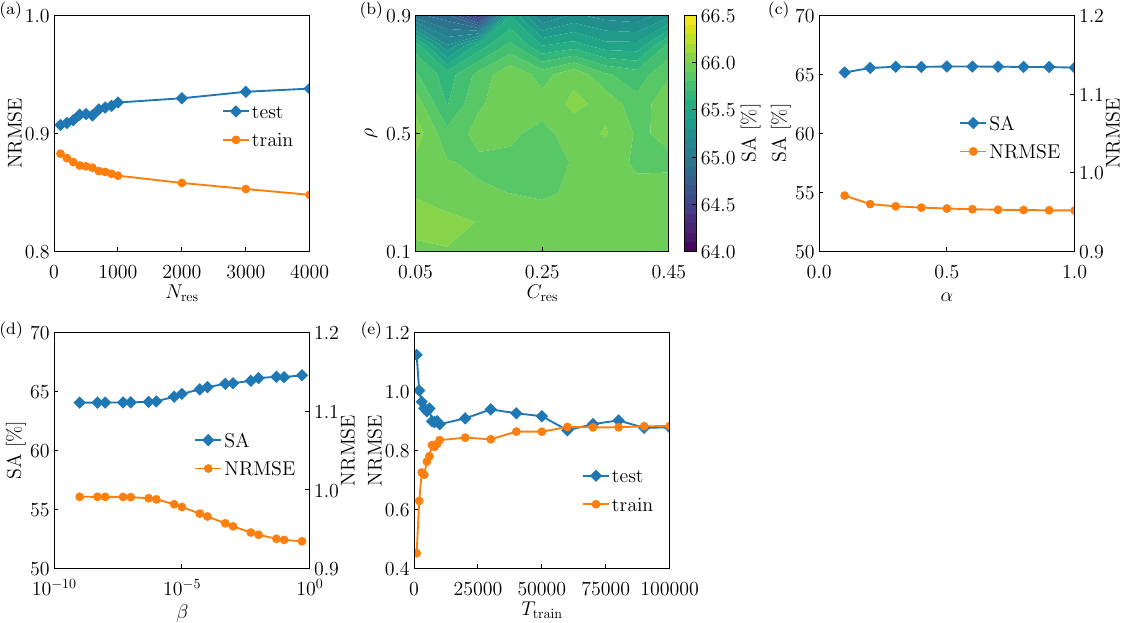}
\caption{
Dependence of hyper-parameters.
(a) Number of reservoir nodes $N_{\mathrm{res}}$,
(b) spectral radius $\rho$ and connectivity $C_{\mathrm{res}}$,
(c) leak ratio $\alpha$,
(d) regularization parameter $\beta$,
and
(e) training length $T_{\mathrm{train}}$.
}
\label{fig:hyperparameter}
\end{figure}

Figure~\ref{fig:hyperparameter}(a)
shows $N_{\mathrm{res}}$-dependence of the training error and the test error,
i.e., NRMSE in the training phase and in the testing phase,
where other five hyper-parameters are fixed to those used in the main text.
The training error decreases as $N_{\mathrm{res}}$ increases.
On the other hand,
the test error increases owing to the overfitting.
The connectivity $C_{\mathrm{res}}$ of $W_{\mathrm{res}}$ and the spectral radius $\rho$ (Fig.~\ref{fig:hyperparameter}(b)),
the leak ratio of LI-ESN $\alpha$ (Fig.~\ref{fig:hyperparameter}(c)),
and the regularization parameter of the ridge regression $\beta$ (Fig.~\ref{fig:hyperparameter}(d))
improve the prediction accuracy no more than 5\%.
Therefore, these four parameters do not affect the result,
and
$\rho = 0.7$, $C_{\mathrm{res}}=0.05$, $\alpha = 0.3$, and $\beta = 5 \times 10^{-4}$ are used in this paper.
The training length $T_{\mathrm{train}}$ does not affect the training error and the testing error
as long as $T_{\mathrm{train}} \geq 1 \times 10^4$ (Fig.~\ref{fig:hyperparameter}(e)),
but small $T_{\mathrm{train}}$ is efficient from the viewpoint of the computational time,
and hence $T_{\mathrm{train}} = 4 \times 10^4$ is used.

\section{Cross Validation}
\label{sec:crossvalidation}

\begin{figure}
\includegraphics{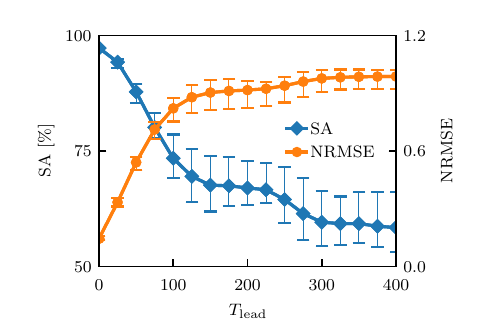}
\caption{
 SA and NRMSE in the random permutation cross validation.
 The error bars represent the minimal and maximal values in the 10 trials.
}
\label{fig:crossvalidation}
\end{figure}

The random permutation (shuffle and split) cross validation~\cite{JMLR:v12:pedregosa11a} is used
to confirm the statistical validity of the prediction.
The data set with time series length $2 \times 10^5$ is split into $10$ intervals.
Three intervals out of the $10$ intervals are randomly chosen.
The first two intervals out of the three intervals and the last one are respectively used
for the training and the testing.
This random procedure is executed 10 times.
The SA and the NRMSE statistically evaluated by the cross validation
are shown in Fig.~\ref{fig:crossvalidation}.
The mean values in the $10$ trials are almost equal to those shown in Fig.~\ref{fig:sanrmse}.
Therefore, the prediction methodology is statistically validated.
As the SA and the NRMSE respectively decreases and increases,
their error bars representing the minimal and maximal values in the 10 trials
become long.
It confirms that the long-term prediction is unsuccessful.
It should be noted that
the lengths of the error bars reach their own saturated values at $T_{\mathrm{lead}} \gtrapprox 125$.

\end{document}